\documentclass[prl,amsmath,amssymb,longbibliography,superscriptaddress,twocolumn]{revtex4-1}
\usepackage[T1]{fontenc}
\usepackage[dvipsnames,rgb,dvips]{xcolor}
\definecolor{matlabgreen}{RGB}{119,172,48}
\definecolor{matlabred}{RGB}{217,83,25}
\definecolor{matlabblue}{RGB}{0,114,189}
\definecolor{matlabpurple}{RGB}{126,47,142}
\definecolor{matlabyellow}{RGB}{237,177,32}
\definecolor{matlablightblue}{RGB}{0,114,189}
\definecolor{cloudgray}{RGB}{229.5,229.5,229.5}

\newcommand{\la}{\left \langle}
\newcommand{\ra}{\right \rangle}
\newcommand{\lp}{ \left(}
\newcommand{\rp}{\right)}
\newcommand\ve[1]{\boldsymbol{#1}}

\newcommand{\das}{\text{Da}_\text{s}}

\newcommand{\dad}{\text{Da}_\text{d}}

\newcommand{\rhow}{\rho_\text{w}}
\newcommand{\md}{\mathrm d}
\newcommand{\LP}{P_{\rm L}}

\usepackage{graphicx}
\usepackage{overpic}
\graphicspath{{Figs/}}

\begin{document}
\title{Lagrangian supersaturation fluctuations at the cloud edge}

\author{J. Fries}
\affiliation{Department of Physics, Gothenburg University, SE-41296 Gothenburg, Sweden}
\author{G. Sardina}    
\affiliation{Department of Mechanics and Maritime Sciences, Chalmers University of Technology, 41296 Gothenburg, Sweden}
\author{G. Svensson}
\affiliation{Department of Meteorology and Bolin Centre for Climate Research, Stockholm University, Stockholm, Sweden}
\affiliation{Department of Engineering Mechanics, KTH Royal Institute of Technology, Stockholm, Sweden}
\author{A. Pumir}
\affiliation{Univ. Lyon, ENS de Lyon, Univ. Claude Bernard, CNRS, Laboratoire de Physique, F-69342,
Lyon, France}
\author{B. Mehlig}
\affiliation{Department of Physics, Gothenburg University, SE-41296 Gothenburg, Sweden}
\begin{abstract}
Evaporation of cloud droplets accelerates when turbulence mixes dry air into the cloud, affecting droplet-size distributions in atmospheric clouds, combustion sprays, and jets of exhaled droplets. The challenge is to model local correlations between droplet numbers, sizes, and supersaturation, which determine supersaturation fluctuations along droplet paths ({\em Lagrangian} fluctuations). We derived a statistical model that accounts for  these correlations. Its predictions are in quantitative agreement with results of direct numerical simulations, and it explains the key mechanisms at play.
\end{abstract}\maketitle

When dry air is mixed into a cloud, water droplets at the cloud edge evaporate. This causes the droplet-size distribution to broaden~\cite{beals2015holographic}, a pre-requisite for
rain formation~\citep{grabowski2013growth}.
Similar processes  occur in combustion sprays~\cite{Sirignano:2014,jenny2012modeling,villermaux2017fine}, and for
respiratory droplets in exhaled jets of air~\cite{Wells,mittal2020flow,Wang:2022,chong2021extended,Bagheri:2023}.
In these systems, an essential physical ingredient is that evaporating droplets saturate the surrounding air. But the
subtle coupling between phase change and turbulent mixing at widely separated turbulent scales \citep{bodenschatz2010can} makes it difficult to predict the local supersaturation and, as a consequence,
droplet-size distributions.
In this work, we study the interplay between these processes and their influence on 
droplet growth, focusing on the parameter regime relevant to the edge of a cloud.

The distribution of supersaturation at droplet positions -- the {\em Lagrangian} distribution -- can be strongly non-Gaussian.
Direct numerical simulations (DNS) of transient mixing of a three-dimensional slab of cloudy 
air with the surrounding dry air  [Fig.~\ref{fig:1}({\bf a})] show exponential tails~\cite{Kumar:2013,kumar2012extreme,kumar2014lagrangian,kumar2018scale}.
Non-Gaussian supersaturation fluctuations are also seen in cloud-chamber experiments~\citep{thomas2021water,prabhakaran2022sources,macmillan2022direct}.
Without phase change, a passive scalar field mixed by turbulence can exhibit non-Gaussian concentration fluctuations  in the presence of a mean scalar gradient~\citep{Pumir+91,warhaft1991probability,Gollub+91}. Without the mean gradient, however, the steady-state distribution  is essentially Gaussian \citep{jayesh1992probability,warhaft2000passive,kumar2012extreme,kumar2014lagrangian}. Non-Gaussian tails may appear in transient mixing~\cite{Villermaux:2019}, but must eventually disappear 
in a homogeneous system.

Whether the tail of the Lagrangian supersaturation distribution is Gaussian or not makes a significant difference, because the tail 
 determines how rapidly certain droplets evaporate, and thereby influences the sizes of the remaining droplets, and thus the droplet-size distibution in unkown ways.
 
 The key question is thus how droplet phase change affects the Lagrangian supersaturation distribution. When phase change is  frequent and rapid, no
consideration of passive-scalar mixing can explain the non-Gaussian relaxation of the Lagrangian supersaturation distribution.
Large-eddy simulations of droplet growth by condensation in a cloud chamber \cite{prabhakaran2022sources} show an anticorrelation between supersaturation $s(\ve x,t)$ and the local droplet-number density $n(\ve x,t)$ in the steady state: in regions with many droplets, the air is strongly subsaturated (very negative values of $s$), while 
it is less so in regions with few droplets (less negative $s$). 
It is plausible that this effect
may change the tails of the Lagrangian supersaturation distribution, but 
existing stochastic models \cite{sardina2015continuous,paoli2009turbulent,siewert2017statistical,chandrakar2016aerosol,abade2017broadening,fries2021key,pinskymixing3,jeffery2007inhomogeneous} 
cannot explain this, because they do not describe how $s(\ve x,t)$ is affected by phase change locally.

We derived a statistical model
that describes transient Lagrangian supersaturation fluctuations developing from an initial inhomogeneity, such as the 
configuration shown in Fig.~\ref{fig:1}({\bf a}), used in earlier DNS studies~\cite{Kumar:2013,kumar2012extreme,kumar2014lagrangian,kumar2018scale,fries2021key}.
We show here that the model captures the key mechanisms that determine the shape of the Lagrangian supersaturation distribution.
First, when mixing occurs on time scales much shorter than phase change, non-Gaussian tails form only during the initial transient, and 
large-time relaxation is characterised by a Gaussian distribution of $s(\ve x,t)$.
 Second, in the opposite limit of rapid phase change, the distribution is no longer Gaussian. Strong phase change drives the mean of the distribution close to its upper bound, while the variance decays more slowly. The distribution is
 squeezed and becomes non-Gaussian. This is reflected in a strong positive correlation
 between $n$ and $s$, formed 
  because phase change tends to establish saturation in regions with many droplets. This mechanism -- the opposite of the effect described in Ref.~\cite{prabhakaran2022sources} -- explains
the tails in the Lagrangian supersaturation distribution described in earlier studies~\cite{kumar2012extreme,Kumar:2013,kumar2014lagrangian,kumar2018scale}. Beyond this qualitative explanation, the model predicts supersaturation distributions that are in excellent quantitative agreement with earlier DNS results. 
The key to success is that the model  inherits its supersaturation dynamics from first principles, rather than imposing an external driving resulting in a Gaussian steady state~\cite{paoli2009turbulent,sardina2015continuous,abade2017broadening}.

We start from  simplified microscopic equations \cite{fries2021key}
governing droplet evaporation in turbulent flow:
\begin{subequations}
\label{eq:mic}
\begin{align}
\label{eq:ns}
&\partial_t\ve u+(\ve u\cdot \ve \nabla)\ve u = -\varrho_{\rm a}^{-1}\ve \nabla p+\nu\nabla^2\ve u\,,\quad \ve \nabla \cdot \ve u=0\,,\\
\label{eq:supsat} &\partial_t s + (\ve u\cdot\ve \nabla) s=\kappa \nabla^2 s - A_2 C_d\,,\\
&\tfrac{{\rm d}}{{\rm d}t} \ve x = \ve u\,, \quad \tfrac{{\rm d}}{{\rm d}t} r = A_3 s/r\,.\label{eq:droplet_r}
\end{align}
\end{subequations}
Eq.~(\ref{eq:ns}) is the Navier-Stokes equation for the incompressible fluid-velocity field $\ve u(\ve x,t)$, where $\varrho_{\rm a}$ is the mass density of air, and $\nu$ its kinematic viscosity.
Eq.~(\ref{eq:supsat}) describes
supersaturation $s = q_{\rm v}/q_{\rm vs}-1$,
with water-vapour mixing ratio $q_{\rm v}=\varrho_{\rm v}/\varrho_{\rm a}$, 
the ratio of the mass densities of
vapour and air, and $\kappa$ is  the diffusivity of supersaturation. Further, $C_d(\ve{x},t)  =\tfrac{4\pi}{3} \rhow n(\ve x,t) \overline{\tfrac{\rm d}{{\rm d}t} r^3}$ 
is the local
rate of change of droplet mass, averaged over all droplets in the vicinity of $\ve x$ with droplet radius $r$.
Here $\rhow$ is the liquid-water density, and $n(\ve x,t)$ is the droplet-number density.
Eqs.~(\ref{eq:droplet_r}) state that the droplets follow the flow, and 
 how the droplet radius $r$ changes \citep{yau1989short}.

We emphasise that 
 the dynamics (\ref{eq:mic}) 
is transient, and tends towards a well-mixed steady state. 
The variance $\sigma_s^2(t)$ of supersaturation fluctuations tends to zero  due to dissipation  (with scalar dissipation rate $\varepsilon_s\equiv 2 \kappa \langle |\ve \nabla s |^2\rangle $) 
and phase change:
\begin{align}\label{eq:var}
\tfrac{\rm d}{{\rm d}t} {\sigma_s^2}= - \varepsilon_s - 2 A_2 \lp\, \langle s  C_d \rangle
-  \langle s \rangle \langle  C_d \rangle\, \rp\,.
\end{align}
Here, the averages are over a given microscopic confiugration at time $t$. 

How the steady state is approached depends on the non-dimensional parameters of the problem.
We non-dimensionalise Eqs.~(\ref{eq:mic},{\ref{eq:var}) as follows~\cite{fries2021key}: time, velocities, and positions with the large-eddy turnover time $\tau_L= k/\varepsilon$ (with turbulent kinetic energy $k$ and kinetic dissipation rate $\varepsilon$), and 
the turbulent r.m.s velocity $u_0=\sqrt{2\,k/3}$; supersaturation with $|s_{\rm e}|$, where $s_{\rm e}$ is the initial subsaturation of the dry air outside the cloud; 
droplet radii with the initial average droplet radius $r_0$.
In the limit of large Reynolds number, the following non-dimensional parameter remain:
the Damk\"ohler numbers $\dad = \tau_L/\tau_{\rm d}$ and $\das = \tau_L/\tau_{\rm s}$
(with supersaturation relaxation time $\tau_{\rm s}$ and droplet-evaporation time $\tau_{\rm d}$ defined as in \cite{fries2021key}), 
and the volume fraction $\chi$ of cloudy air [Fig.~\ref{fig:1}({\bf a})].
The Schmidt number Sc$=\nu/\kappa$ is of order unity~\cite{Vaillancourt:2001}.

Earlier attempts to analyse the process were based on statistical models of mixing and evaporation that describe droplet evaporating in direct response to a
 spatially inhomogeneous mean field given by an ensemble average, $\la s(\ve x,t) \ra$~\cite{jeffery2007inhomogeneous,pinskymixing3,pinskyplume1}. A more sophisticated model  \cite{fries2021key}
accounts for how droplet-phase change is affected by Lagrangian supersaturation fluctuations, but still assumes
 that they decay exponentially towards the mean. 
Both types of models 
explain how the extent of complete droplet evaporation depends on the Damk\"ohler numbers, but fail to reproduce the far tail of cloud-droplet size distribution obtained in DNS 
\cite{kumar2012extreme,Kumar:2013,kumar2014lagrangian,kumar2018scale}. A likely reason is that these models underestimate  the magnitude of supersaturation fluctuations.
A further shortcoming is that these models assume exponential relaxation of supersaturation to $\langle s(\ve x,t)\rangle$. As a consequence, they fail to reproduce the passive-scalar limit~\cite{pope2000turbulent},
namely Gaussian supersaturation fluctuations with exponentially decaying variance.

The question of how a passive-scalar distribution relaxes as it is mixed by turbulence has a long history.
Eswaran \& Pope \cite{eswaran1988direct} analysed this process systematically using DNS. Their results inspired and benchmarked increasingly accurate
 models for passive-scalar mixing~\cite{valino1991binomial,fox1992fokker,fox1995spectral,meyer2006mixing,chen1989probability,pope1991mapping,Villermaux:2019}. 

 {\em Model.} The mapping-closure approximation \cite{chen1989probability,pope1991mapping} 
describes how the shape of a passive-scalar distribution changes as the scalar is mixed in turbulence. The approximation relies only on one-point statistics. Correlations are not needed. As a consequence, the approximation does not predict the speed of the mixing process, but yields accurate and robust predictions for the sequence of shapes of the distribution. Therefore it is ideally suited for our purposes. 
The mapping closure for 
 {\em Eulerian} supersaturation fluctuations 
 starts from 
\begin{align}\label{eq:ansats}
s(\ve x,t) = X[\xi(\ve x/\lambda(t)),t]\,,
\end{align}
 where $\lambda(t)$ a time-dependent length scale, 
 $\xi$ is a spatially smooth
random Gaussian field with mean zero and unit variance, and
 $X$ is  the time-dependent mapping from $\xi(\ve x/\lambda(t))$ to $s(\ve x,t)$.
 Inserting (\ref{eq:ansats}) into (\ref{eq:supsat}), one obtains
\begin{equation}\label{mapevol}
\partial_tX= \varphi(t)\big(\!-\!\eta {\partial_\eta X}
+{\partial_\eta^2X}\big)\!-\!\das \la C_d| s \!=\! \!X\ra \,.
\end{equation}
In comparison to the original model~\cite{chen1989probability,pope1991mapping}, Eq.~(\ref{mapevol})
contains the phase-change term
$\langle C_d| s \!=\! \! X\rangle\!=\! X \langle r N/V|s \!=\! \!X\rangle$, with droplet number $N$ and spatial volume $V$.
We approximate this term by a mean-field decoupling of the conditional average (see SM~\cite{sm} for details)
\begin{equation}
\label{eq:cd_cond}
\langle C_d| s \!=\! \! X\rangle =  X
\langle r|s\!=\! \!X\rangle \langle N|s \!=\! \!X\rangle\!/\!\langle V|s \!= \!\!X\rangle\,.
\end{equation}
\begin{figure}
\centering
\begin{overpic}[width=\columnwidth]{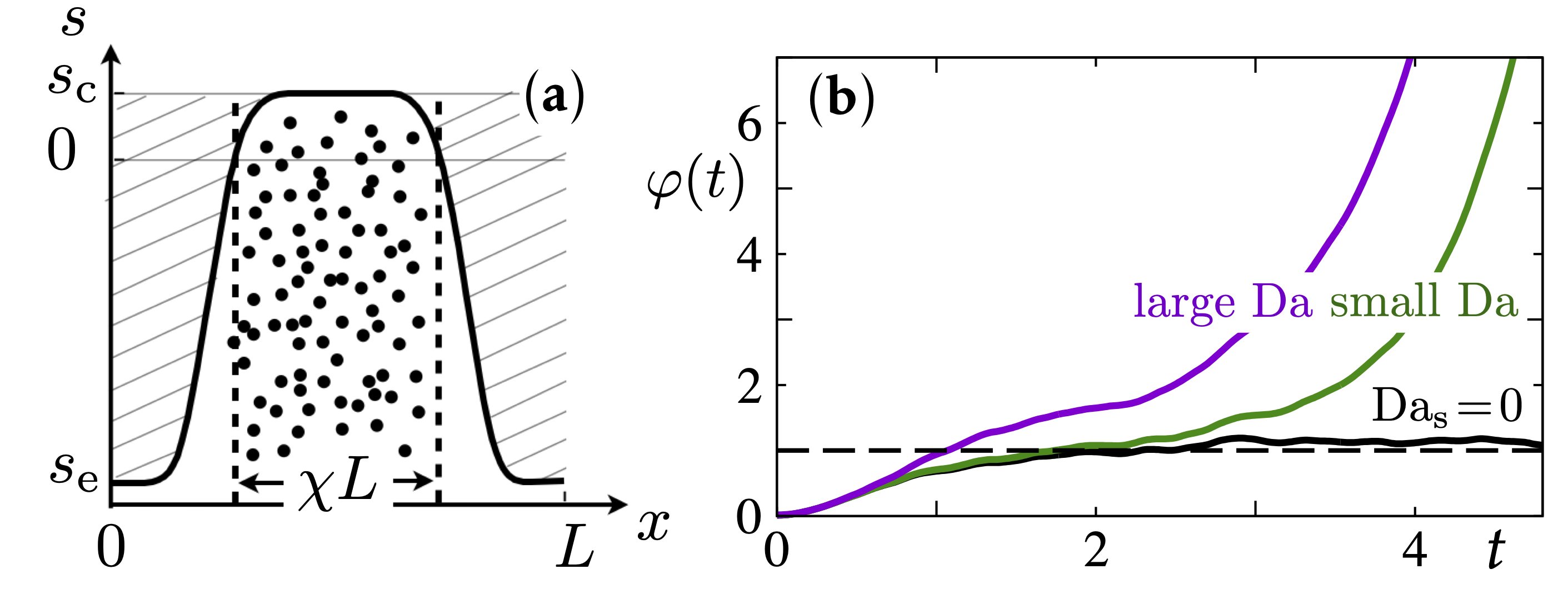}
\end{overpic}
\caption{\label{fig:1} 
({\bf a})~Initial condition used 
in~\cite{kumar2012extreme,Kumar:2013,kumar2014lagrangian,kumar2018scale,fries2021key}, a three-dimensional slab of cloudy air containing droplets with initial number density~$n_0$ ($\bullet$)
and supersaturation  $s_{\rm c} \geq 0$, surrounded by dry subsaturated air with $s_{\rm e} < 0$  (hashed). The supersaturation profile is shown as a solid line. 
The cloudy air occupies a volume fraction $\chi$.
({\bf b}) Statistical-model rate $\varphi(t)$ as a function of time for different
Damk\"ohler numbers, see Eq.~(\ref{mapevol}) and SM~\cite{sm} for details. The dashed line is the steady-state limit $\varphi_\ast=C_\phi/2$ for a passive scalar
with $C_\phi=2$~\citep{pope2000turbulent} (see text).  Parameters from Ref.~\cite{kumar2012extreme}: $\chi=0.4$, $\das = 0.80, \dad = 0.073$ (small Da), and $\das = 8.0, \dad = 0.73$ (large Da). The value  of $\tau_L$ in our DNS differs slightly from that in  Ref.~\cite{kumar2012extreme} due to statistical variability in the forcing.}
\end{figure}

The factor $\varphi(t) = \kappa/[u_0^2 \tau_L\lambda^2(t)]$
in Eq.~(\ref{mapevol}) is the non-dimensional relaxation rate
of the Eulerian distribution. 
How $\varphi(t)$ changes as a function of time is determined by processes at both small and large length scales, as the following argument shows.
For  passive-scalar mixing, 
the scalar variance decays exponentially in the self-similar regime~\citep{pope2000turbulent}, $\tfrac{\rm d}{{\rm d}t}\sigma_{\rm s}^2 =-C_\phi  \sigma_{\rm s}^2$.
 with $C_\phi \approx 2$.
In this case, $\varphi(t)$ approaches the steady-state value,  $\varphi_\ast \sim C_\phi/2$.
The steady state emerges as a  balance between the scalar variance cascading towards large wave numbers and  rapid dissipation at large wave numbers. In  physical dimensions, the steady-state length scale
$\lambda_\ast = [2\kappa k/(C_\phi\varepsilon)]^{1/2}$ equals $\lambda_{\rm T}(5C_\phi {\rm Sc})^{-1/2}$ where 
$\lambda_{\rm T} =(10\nu k/\varepsilon)^{1/2}
$ is the Taylor microscale. 
In other words, both large-scale mixing  and small-scale diffusion matter.

With phase change, $\varphi(t)$ is unknown. We determine it using DNS, 
see Supplemental Information (SI)~\cite{sm} for details.
The results are summarised in Fig.~\ref{fig:1}({\bf b}) which shows how $\varphi(t)$
evolves as a function of $t$.
For passive-scalar mixing, the predicted plateau at
$\varphi_\ast$ is approached after two large-eddy turnover times, at  $t\approx 2$.
With phase change, $\varphi(t)$ is
larger, corresponding to smaller $\lambda(t)$.
This is consistent with the notion that phase change generates supersaturation gradients by driving the air towards saturation where droplets exist, 
while subsaturated regions without droplets remain subsaturated.

To obtain the {\em Lagrangian} supersaturation fluctuations, 
\citet{pope1991mapping} suggested to use a Langevin equation for~$\xi(t)$
$ \md \xi  = -R(t) \xi \md t + [2 R(t) ]^\frac{1}{2} \md v\,,$
where  $\md v$ is the increment of a Gaussian random process, 
and to compute the supersaturation as $s(t) = X[\xi(t),t]$. 
The Langevin equation ensures that the distribution of $\xi(t)$ relaxes to a normalised Gaussian, and the function $X(\eta,t)$ maps this Gaussian to the Eulerian supersaturation distribution. This ensures that the Lagrangian supersaturation distribution relaxes to the Eulerian one, as required.  
We set $R(t) = C \varphi(t)$, 
where $C$ is a constant.
This is motivated -- at least for a passive scalar -- by the fact that supersaturation fluctuations due to turbulent mixing experienced by a fluid element reflect the diffusive term $\kappa \nabla^2 s$ in Eq.~(\ref{eq:supsat}), and that the fluctuations of this term are proportional to  $\varphi(t)$ under the mapping closure.
Comparison with DNS  shows that 
$R(t) = C \varphi(t)$  works very well for the first two large-eddy turnover times, 
for $\das$ up to $8.0$. We find that $C$ decreases as $\das$ increases (see SI~\cite{sm}), because
phase change tends to maintain saturation in  regions with droplets. 
For  times much larger than the large-eddy turnover time, the precise form of $R(t)$  does not matter  because
the Lagrangian distribution has almost relaxed to the Eulerian one. 

\begin{figure}
\centering
\begin{overpic}[width=\columnwidth]{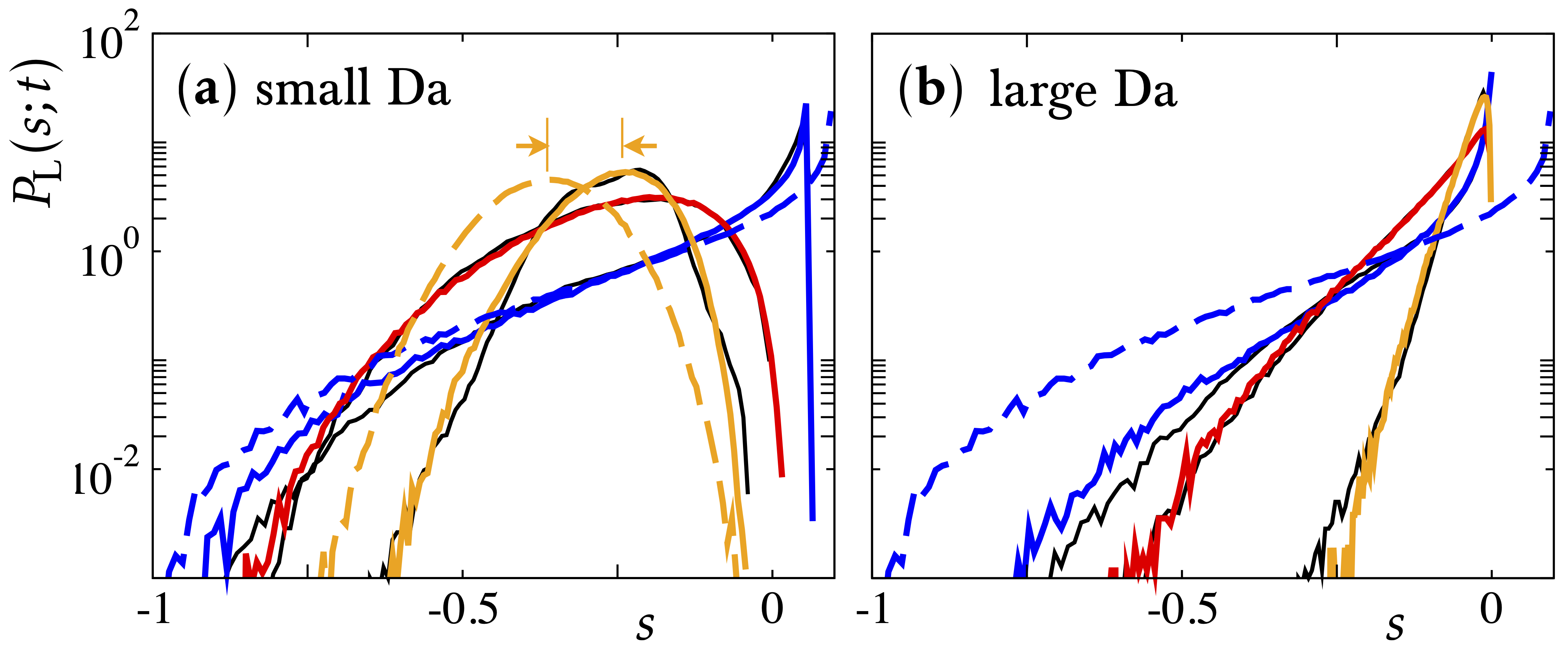}
\end{overpic}
\caption{\label{fig:2} Lagrangian supersaturation distributions.
({\bf a}) 
$\das = 0.80$, $\dad = 0.073$, and $\chi = 0.4$ (parameters from Ref.~\cite{kumar2012extreme}). DNS results (see SI~\cite{sm} for details):
solid black  lines. 
Statistical-model simulations: solid colored lines ($t=0.68$, blue; $t=1.69$, red; $t=2.36$, orange).
Statistical-model simulations for a passive scalar  ($t=0.68$, dashed blue; $t=2.36$, dashed orange).  The shift due to phase change is indicated by horizontal arrows.
Panel ({\bf b}): same as in ({\bf a}), but for $\das = 8.0$, $\dad=0.73$ (parameters from Ref.~\cite{kumar2012extreme}). The DNS results shown here were obtained using the same turbulent velocity field, and the same initial droplet configuration.}
\end{figure}
{\em Results.} 
Fig.~\ref{fig:2} shows the Lagrangian supersaturation distribution $\LP(s;t)$  from  the statistical model (solid lines), compared
with earlier DNS results \cite{kumar2012extreme} (dashed lines). 
Shown are two cases: small and large Damk\"ohler numbers. We see that the statistical model reproduces the  DNS results quantitatively.

For small $\das$, the effect of phase change is small at short times, the distribution is close to that of a passive scalar (blue dashed lines).
Supersaturation behaves  essentially like a passive scalar during the first large-eddy turnover time, $t\sim 1$.
At later times, droplet-phase change matters more, but its effect is straightforward, it causes the peak of the distribution to shift somewhat compared to the distribution for $\das=0$, to less negative values of $s$, while supersaturation fluctuations are still approximately Gaussian [Fig.~\ref{fig:2}({\bf a})].  
For larger values of $\das$, by contrast, the evolution of the supersaturation distributions looks very different [Fig.~\ref{fig:2}({\bf b})].
The distribution remains non-Gaussian  at large times.

To pin down the precise mechanism, 
we followed the droplet-number density $n(t)$  and  the local supersaturation $s(t)$ for different fluid parcels in DNS. 
The results are summarised in Fig.~\ref{fig:3} which shows the conditional average
of the local-droplet number density conditional on the surrounding supersaturation for the same parameters as in Fig.~\ref{fig:2}. 
For small Damk\"ohler numbers, the average does not change much during the time shown, it is still strongly influenced by the initial condition.  
For large $\das$, by contrast, the average changes 
rapidly, and the positive correlation between $n(t)$ and $s(t)$ increases significantly.
The statistical 
model captures this very well. 
The mechanism is simply that 
a parcel containing many droplets cannot remain sub- or supersaturated for long, because
phase change  drives the air quickly towards saturation when $\das$ is large. As a consequence, parcels with few droplets tend to have much more negative
values of $s$, compared with 
a parcel with small $\das$. 
The strong suppression of the conditional average $\langle n(t)|s(t)\!=\!s\rangle$ at large $\das$ explains how the non-Gaussian tails evolve in Fig.~\ref{fig:2}({\bf b}):
the left tail of $\LP(s;t)$ disappears quickly as time increases, because droplets saturate their surroundings, and therefore fewer of them experience very dry air.
 
For small $\das$‚ the mapping closure results in Gaussian relaxation. Phase change
causes  non-Gaussian tails. In order to describe these tails, it is
necessary  to condition $C_{\rm d}$ on supersaturation, Eq.~(\ref{eq:cd_cond}). 
The conditioning also ensures that the supersaturation fluctuations remain bounded, as they must because neither phase change 
nor mixing can turn subsaturated into supersaturated air. 

\begin{figure}
\centering
\begin{overpic}[width=\columnwidth]{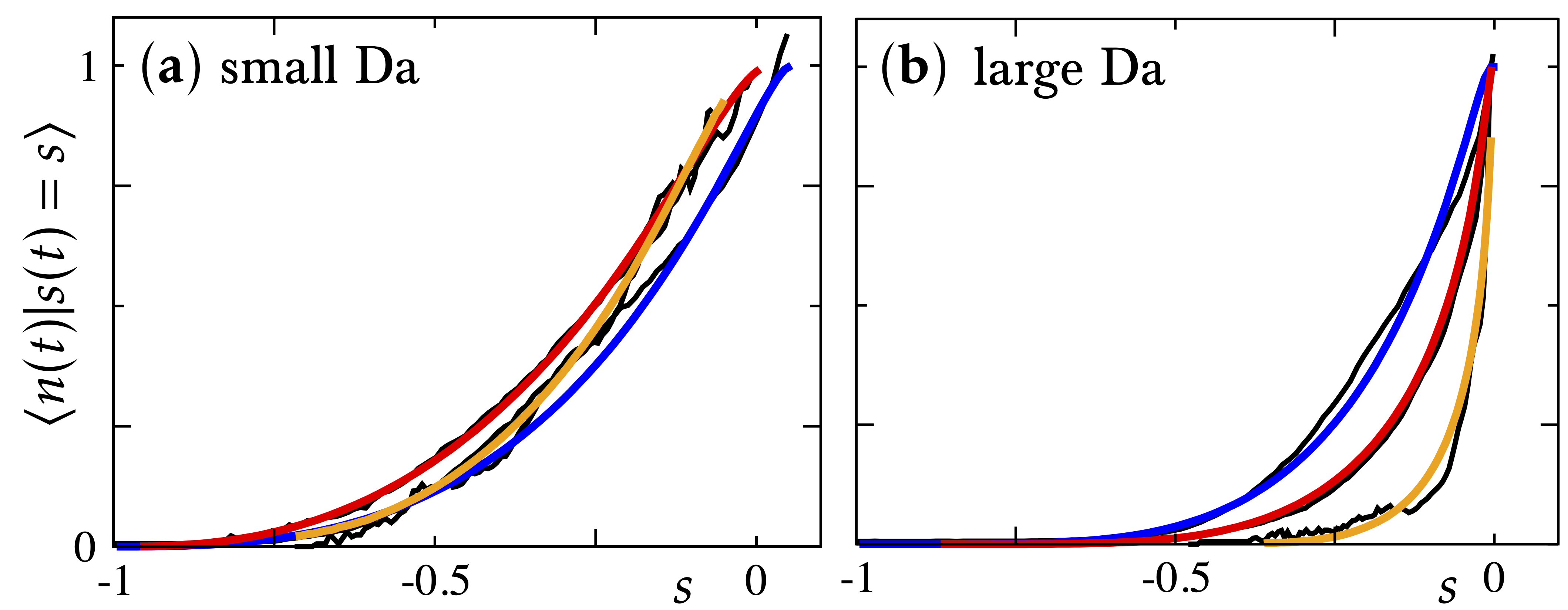}
\end{overpic}
\caption{\label{fig:3} Correlations between  droplet-number density and  supersaturation. Average $\langle n(t)|s(t)=s\rangle$ of droplet-number density at time $t$ conditional on local supersaturation~$s$. 
 ({\bf a})~small Damk\"ohler numbers, $\das = 0.80$, $\dad = 0.073$, and $\chi = 0.4$.
Statistical-model simulations: solid colored lines ($t=0.68$, blue; $t=1.69$, red; $t=2.36$, orange). DNS results:  solid black lines lines. 
 ({\bf b}): same as in ({\bf a}), but for $\das = 8.0$ and $\dad=0.73$ (parameters from \citet{kumar2012extreme}). 
 }
\end{figure}

{\em Discussion.}
We begin by discussing in more detail, how our results
relate passive-scalar mixing.
Eswaran and Pope~\cite{eswaran1988direct} described the  shape change of the Eulerian passive-scalar distribution as a function of time. The initial condition [Fig.~\ref{fig:1}({\bf a})]
dictates that the Eulerian distribution is, initially, the sum of two  narrow peaks, located at $s=s_{\rm c}$ and $s_{\rm e}$. It relaxes first to a U-shaped form. The left tail 
of the Lagrangian supersaturation distribution reflects how the  Eulerian peak at  $s=s_{\rm c}$ broadens. 
At large times, the Eulerian U-shaped distribution relaxes to a Gaussian. 

Our model predicts 
the same for small but not negligible $\das$, with one important difference: phase change causes the mean of the Lagrangian supersaturation distribution to shift to the right [Fig.~\ref{fig:2}({\bf a})], while
mixing causes the distribution to narrow, remaining approximately Gaussian. In this case,
the mapping is approximately given by
$X(\eta,t) = \sigma_s(t) \eta + \mu_s(t)$.
Inserting this into Eq.~(\ref{mapevol}) and assuming
passive-scalar relaxation of the width,
${\rm d}\sigma_s/{\rm d}t = -C_\phi\sigma_s/2$,
yields
$ {\rm d}\mu_s/{\rm d}t = -\das \mu_s$. 
So the standard deviation decays
more rapidly than the mean for small $\das$,
consistent with Gaussian relaxation.

At large $\das$, the time evolution of the Lagrangian supersaturation distribution is strongly affected by phase change, resulting in persistent non-Gaussian tails  [Fig.~\ref{fig:2}({\bf b})].
Our model explains why the Lagrangian supersaturation distributions relax so differently for small and large $\das$.
Rapid phase change quickly drives the mean of the distribution towards the upper bound of the supersaturation distribution.
As a result, the distribution is squeezed towards $s=0$,
thus preventing a Gaussian from forming. 
The distribution is bounded because subsaturated air can not obtain a positive supersaturation through droplet evaporation. 
Therefore saturation ($s=0$) constitutes an upper bound for the Lagrangian supersaturation fluctuations. 

We contrast our results with those of 
 \citet{prabhakaran2022sources}. They found a negative correlation between $n(t)$ and $s(t)$ in LES designed to model droplet condensation in a cloud chamber. 
 Their system is statistically stationary,  but the statistical model highlights the mechanism leading to their findings. In their case, the air is saturated or supersaturated, so droplets tend to grow by condensation. The resulting drive towards saturation gives rise to a positive  correlation between $n(t)$ and $s(t)$.

The model describes not only the Lagrangian supersaturation fluctuations quantitatively, but also the droplet-size distribution (not shown, see SI~\cite{sm}). Our earlier model \cite{fries2021key} yielded qualitative but not quantitative agreement, highlighting the importance of correlations between $n$ and $s$.

We recall that Damk\"ohler numbers in atmospheric clouds tend to be large, simply because the relevant length scale $L$ is large, causing large $\tau_{\rm L}$.  It is tempting to argue that the persistent left tail  of the Lagrangian supersaturation distribution at large Damköhler numbers in Fig.~\ref{fig:2}({\bf b}) is more representative of atmospheric relaxation than the Gaussian relaxation at small Damk\"ohler numbers. However, 
the local supersaturation field around individual droplets is hard to observe {\em in situ}. Observations resolving supersaturation at
larger scales, of the order of one metre, indicate Gaussian distributions \citep{siebert2017supersaturation}, but better resolved laboratory measurements reveal skewed distributions \citep{anderson2021effects}, as predicted by our model.

Here we analysed moist systems, with $\dad/\das  \propto (\rhow n_0 r_0^3)^{-1} \sim 0.1$. We expect the present model to apply equally well to dry clouds where complete droplet evaporation occurs frequently, but we have not yet explored this regime.

\citet{villermaux2017fine} measured the joint dynamics of vapour and droplets in a dense acetone spray. 
They analysed vapour concentrations for different flow configurations considering
the limit of large droplet-number density $n_0$ and large Damk\"ohler numbers,  where the droplets in the spray prevent each other from evaporating, but evaporate instantenously in dry air. In this limit, correlations
between $n(\ve x,t)$ and $s(\ve x,t)$
are extreme, and it remains to be seem whether they can be captured by our model.
More generally, it is of interest to compare the accuracy of the present mapping-closure model with predictions of the linear-eddy model~\cite{kerstein1988linear}, 
where turbulent stretching and folding is represented by a one-dimensional map~\cite{krueger1993linear,su1998linear,hoffmann2019inhomogeneous}. 

{\em Conclusions.} }
We derived a statistical model
for the transient supersaturation fluctuations around droplets near the cloud edge, where turbulence mixes  dry with cloudy air, causing the droplets to evaporate. 
The model explains the key mechanisms determining Lagrangian supersaturation fluctuations, 
and its predictions are in 
quantitative agreement with earlier 
DNS studies of droplet evaporation at the cloud edge~\cite{Kumar:2013,kumar2012extreme,kumar2014lagrangian,kumar2018scale,fries2021key}.
This advance became possible because
the model describes supersaturation dynamics and the local coupling  due to phase change using a  mapping closure, which is known to yield quantitative results for passive-scalar mixing. 
At the same time, the model is simple enough so that it can be used to resolve sub-grid scale effects in large-eddy simulations with high precision.

We stress that the present model, unlike earlier statistical models, accounts
for local correlations between droplet numbers and supersaturation. This opens the possibility to model the dynamics of denser turbulent aerosols, such as industrial sprays.

\begin{acknowledgments}
JF was supported by grants from  the Knut and Alice Wallenberg (KAW) Foundation (no. 2014.0048) and Vetenskapsr\aa{}det  (VR), no.~2021-4452. G. Sardina
acknowledges support from VR (grant no.~2022-03939). 
The computations were enabled by resources provided by the National Academic Infrastructure for Supercomputing in Sweden (NAISS) and the Swedish National Infrastructure for Computing (SNIC) at NSC and PDC partially funded by the Swedish Research Council through grant agreements no. 2022-06725 and no. 2018-05973. 
The collaboration of AP and BM during the  {\em Multiphase~22} program  at the KITP was supported  in part by the National Science Foundation under grant no. NSF PHY-1748958.
\end{acknowledgments}


%
\end{document}


\title{Supplemental Information for\\ \lq Lagrangian supersaturation fluctuations at the cloud edge\rq{}}

\author{J. Fries}
\affiliation{Department of Physics, Gothenburg University, SE-41296 Gothenburg, Sweden}
\author{G. Sardina}    
\affiliation{Department of Mechanics and Maritime Sciences, Chalmers University of Technology, 41296 Gothenburg, Sweden}
\author{G. Svensson}
\affiliation{Department of Meteorology and Bolin Centre for Climate Research, Stockholm University, Stockholm, Sweden}
\affiliation{Department of Engineering Mechanics, KTH Royal Institute of Technology, Stockholm, Sweden}
\author{A. Pumir}
\affiliation{Univ. Lyon, ENS de Lyon, Univ. Claude Bernard, CNRS, Laboratoire de Physique, F-69342,
Lyon, France}
\author{B. Mehlig}
\affiliation{Department of Physics, Gothenburg University, SE-41296 Gothenburg, Sweden}

\maketitle

\section{Description of direct numerical simulations} 
\label{sec:dns}

In this Section, we describe the direct numerical simulations (DNS) used to obtain the data shown in Figs~1({\bf b}), 2, and 3.
The governing equations~(1) were solved in an Eulerian-Lagrangian approach: the Eulerian momentum and supersaturation equations [Eqs.~(1a,b)] were solved using a pseudo-spectral DNS code coupled with the equation of motion for the droplets [Eq.~(1c)]. The numerical solver has been used in previous works on cloud droplet evaporations in turbulence \cite{sardina2015continuous,sardina2018broadening}.  The Eulerian equations were solved in Fourier space with the nonlinear terms calculated in physical space using a standard 2/3 rule for de-aliasing. Direct and inverse fast Fourier transforms were employed. The equations were advanced in time with a low-storage third-order Runge-Kutta method with the diffusive terms analytically calculated, while an Adam-Bashforth scheme was  employed for the nonlinear terms.
Likewise, a low-storage third-order Runge-Kutta temporal scheme was used to  integrate the  equations of motion for the droplets.

 Air velocity and supersaturation at the droplet positions were calculated with a linear interpolation scheme. Linear extrapolation was used to evaluate the $C_d$ on the Eulerian grid. The simulation configurations and parameters were from \citet{kumar2012extreme}. More precisely, we used the parameters given in the
 middle line and the third line of their Table~2. Following  \citet{kumar2012extreme}, the simulation domain was taken to be a cube with a length of $L_x=L_y=L_z=25.6$ {\rm cm}, where homogeneous turbulence was forced to obtain a mean turbulent kinetic dissipation rate of $\varepsilon=\nu \langle |\ve \nabla {\ve u} |^2\rangle =33.75$ ${\rm {c}m}^2\,{\rm s}^{-3}$, or, equivalently, a Kolmogorov scale  of $\eta=(\nu^3/\varepsilon)^{1/4}=1 \,{\rm  mm}$. The turbulent kinetic energy evaluated to $k=3 u_0^2/2=49$ cm$^2$ s$^{-2}$, where $u_0=u_{x,{\rm rms}}$ cm s$^{-1}$ is the turbulent r.m.s. velocity.  We employed $512^3$ collocation points to solve the equation on the Eulerian grid. 
The  turbulence was  forced to maintain a statistically steady state. In Fourier space, the forcing was
 $ \ve{\hat  f}(\ve k,t)=\varepsilon \delta_{\ve k,\ve k_f} \hat{\ve u}(\ve k,t)/\sum_{\ve k_f} |\hat {\ve u}(\ve k_f,t)|^2$, where $\delta_{\ve k,\ve k'}$ is the Kronecker delta, $ \hat{\ve u}(\ve k,t)$ is the Fourier transform of the velocity field, and  the wave vectors are of the form $\ve k_f=[\pm 2\pi/L_x,\pm 2\pi/L_y,\pm 4\pi/L_z]$, as well as  all permutations of the components.

All  simulations had an initial droplet radius $r_0=20 \mu {\rm m}$.
 Initial supersaturation and droplet positions were chosen as illustrated in 
 Fig.~1({\bf a}). In particular, the cloud slab  had an initial droplet-number density of $n_0 = 164$ {\rm cm}$^{-3}$, supersaturation $s_{\rm c}=0.02$, and volume fraction $\chi=0.4$, while the dry region  contained no particles initially, and had supersaturation $s_{e}=-0.2$. Other relevant parameters of the two DNS  runs were: supersaturation Schmidt number $\nu/\kappa=0.7$, $A_2=1/(\airdensity q_{\rm vs})=265$~m$^3$/kg
 ($\airdensity$ is the mass density of dry air, and $q_{\rm vs}$ is the saturation value of the vapor mixing ratio at 270 K), and $A_3=50.7 \mu {\rm m}^2/{\rm s}^{-3}$ in Fig.~2{\bf a}).
 Following \citet{kumar2012extreme}, we increased $A_3$ by a factor of ten for Fig.~2({\bf b}), in order to simulate inhomogeneous mixing conditions.
The values of the Damk\"ohler numbers are summarised in Table~\ref{tab:C}.
  
\section{Details of mapping closure}
\label{sec:mpc}
In this Section, we derive the mapping closure with phase change, Eq.~(4),
and explain how we closed the phase-change term. 
 The description of the mapping closure summarised below differs from  Ref.~\cite{pope1991mapping} only in details. The main difference
is that we assume that the physical supersaturation field
is given by $s(\ve x,t) = X[\xi(\ve x/\lambda(t)),t]$ [Eq.~(3)], rather than working with a surrogate field. 
 
We start from the evolution equation for the 
Eulerian cumulative distribution function $F(S;t)=\mbox{Prob}( s(\ve x,t) < S)$. 
From Eq.~(1b), one obtains an equation for $F$:
\begin{align}\label{cdfevol}
\pder{F(S;t)}{t} = - \la \kappa \nabla^2 s | S \ra\pder{F(S;t)}{S} + A_2 \la C_d | S \ra\pder{F(S;t)}{S}\,.
\end{align}
Here and in the remainder of this Section, we use dimensional units, as in Eq.~(1).
In order to derive Eq.~(4), we need to insert the ansatz $s(\ve x,t) = X[\xi(\ve x/\lambda(t)),t]$, as described by Pope~\cite{pope1991mapping}.
We assume that $X(\eta,t)$ is a non-decreasing function of $\eta$. Now recall that $\xi$ is a standardised Gaussian field. This implies
\begin{align}\label{gausscdf}
F[X(\eta,t);t] = G(\eta)\,,
\end{align}
where $G(\eta)$ is the probability density of a standardised Gaussian random variable.
Differentiating this relation with respect to time yields
\begin{align}\label{cdfevolmapping}
\pder{F(S;t)}{t} = - \pder{X}{t} \pder{F(S;t)}{S}\,.
\end{align}
 Comparing the terms in Eqs.~(\ref{cdfevol}) and (\ref{cdfevolmapping}), one finds:
\begin{align}\label{mappingevol}
\pder{X}{t} = \la \kappa \nabla^2 s | X \ra - A_2 \la C_d | X \ra
\end{align}
To arrive at Eq.~(4), one needs to evaluate the average $\la \kappa \nabla^2 s | X \ra$. The gradients of supersaturation are computed using
the ansatz $s(\ve x,t) = X[\xi(\ve x/\lambda(t)),t]$. This yields:
\begin{subequations}
\begin{align}\label{lapl}
\nabla^2 s = \pderder{s}{x_i}{x_i} = \frac{1}{\lambda^2(t)} \pder{X}{\eta}\pderder{\xi (\ve z)}{z_i}{z_i} + \frac{1}{\lambda^2(t)}   \pder{^2 X}{\eta^2}  \pder{\xi (\ve z)}{z_i} \pder{\xi (\ve z)}{z_i}.
\end{align}
\end{subequations}
Here, we introduced the components $z_i$ of $\ve z  = \ve x/\lambda(t)$, the argument of the Gaussian field $\xi$. Furthermore, summation
 over repeated indices is implied, also in the following. Now we take the average conditional on $\xi=\eta$. Since $\xi$ is a standardised Gaussian field, one has~\cite{pope1991mapping}
\begin{align}
\label{avgg}
\la \pderder{\xi (\ve z)}{z_i}{z_i} \vline \; \xi = \eta \ra = - \frac{\eta}{\Lambda^2}
\quad\mbox{and}\quad
\la  \pder{\xi (\ve z)}{z_i} \pder{\xi (\ve z)}{z_i} \vline \;  \xi = \eta \ra = \frac{1}{\Lambda^2}\quad\mbox{with}\quad 
\Lambda = \la \pder{\xi}{z_i} \pder{\xi}{z_i} \ra^{-\frac{1}{2}}\,.
\end{align}
Averaging Eq.~(\ref{mappingevol}) with (\ref{avgg}) yields Eq.~(4), if we set $\Lambda=1$. This can be done without loss of generality, 
because for general $\Lambda$, the prefactor $\kappa/ \lambda^2(t)$ in the first term on the right-hand side of Eq.~(4) is  replaced by
 $\kappa/[\Lambda \lambda(t)]^2$. Since $\lambda(t)$ is an undetermined function (see next Section), this does not make any difference. 
 We note that~\citet{pope1991mapping} does not assume $\Lambda = 1$. As a consequence, a length scale $\lambda_\theta$ appears in the evolution equations for the mapping $X(\eta,t)$ in his  formulation. Our length scale $\lambda(t)$ equals the product of $\lambda_\theta$ and their time-dependent non-dimensional function $J(t)$, 
 namely $\lambda(t) = J(t)\lambda_\theta$.

\begin{figure}
\centering
\begin{overpic}[width=50mm]{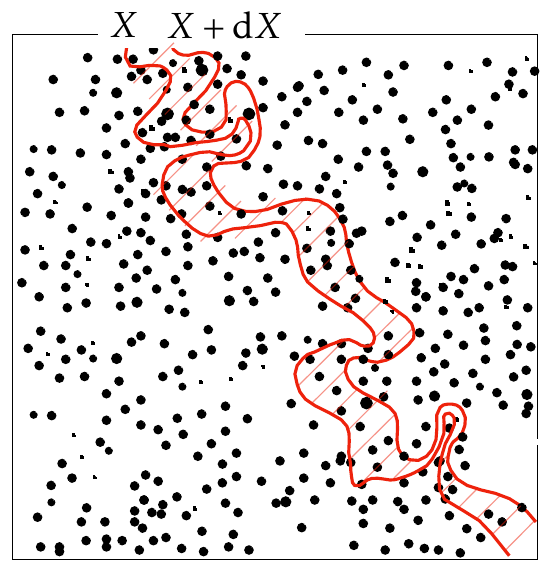}
\end{overpic}
\caption{\label{fig:contours}
Schematic. Shows contours of supersaturation $s(\ve x,t)=X$ and $s(\ve x,t)=X+{\rm d}X$.
 To evaluate the averages in Eq.~(\ref{eq:avgs}), 
 one records the droplet number $N$ and  their sizes
 $r$ in the hashed region of volume (here area) $V$. }
\end{figure}

 We now discuss how we approximated the conditional
 average $\langle C_d|s=X\rangle$ in Eq.~(4). 
 It involves the average
 $\langle r n|s=X\rangle$ which is approximated as
 \begin{equation}
 \label{eq:avgs}
\langle r n|s=X\rangle\approx 
 \langle r|s\!=\! \!X\rangle \langle N|s \!=\! \!X\rangle\!/\!\langle V|s \!= \!\!X\rangle\,,
     \end{equation}
in terms of the average number $N$ of droplets with supersaturation in the interval $[X,X+{\rm d}X]$,  
their sizes, and the volume of regions with that supersaturation (Fig.~\ref{fig:contours}).
The ratio $\langle N|s \!=\! \!X\rangle\!/\!\langle V|s \!= \!\!X\rangle$ can be expressed as
\begin{align}\label{ratiorelation}
\frac{\langle N|s \!=\! \!X\rangle}{\langle V|s \!= \!\!X\rangle} = \frac{N_0}{V}\frac{P_L(X;t)}{f(X;t)}\,.
\end{align}
Here $n_0$ is the droplet-number density, and
\begin{align}
f(X;t) = \pder{}{X}F(X;t)
\end{align}
is the Eulerian probability density of supersaturation~\cite{fries2022mixing}.
The average radius $\langle r |s=X\rangle$ conditional on supersaturation $s = X$ is given by
    \begin{align}
    \label{eq:avgr}
\langle r|s\!=\! \!X\rangle = \int\! \md r \,r \frac{P_L(r,s;t)}{P_L(s;t)} \,.
  \end{align}
  Here $P_L(s;t)$ is the Lagrangian supersaturation distribution discussed in the main text, and $P_L(r,s;t)$ is the corresponding joint distribution of droplet radii and supersaturation.

Following \cite{pope2000turbulent}, we use a Monte-Carlo method to simulate the statistical model, using an ensemble of Lagrangian fluid elements. With each such element there
is an associated value of $\xi(t)$ from which its supersaturation $s(t) = X[\xi(t),t]$ is computed using the mapping $X[\eta,t]$. 
 There are two types of fluid elements,
 one that is used to sample the Eulerian distribution $f(X;t)$,  and the other to sample the Lagrangian distribution $P_L(s,r;t)$ and the corresponding marginal distribution $P_L(X;t)$. Each fluid element  of the latter type is associated with a radius $r(t)$, in addition to supersaturations $s(t)$. In practice, $f(X;t)$, $P_L(X;t)$ in Eq.~(\ref{ratiorelation}),  and the average $\la r|s=X\ra$ in Eq.~(\ref{eq:avgr}) 
 are evaluated by binning the continuous variable $X$. Note that the bin
 $[X,X+{\rm d}X]$ corresponds represent a rather complex shape in configuration space, as schematically illustrated in Fig.~\ref{fig:contours}.
 
\begin{figure}
\centering
\begin{overpic}[width=50mm]{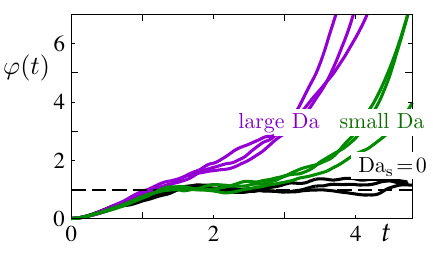}
\end{overpic}
\caption{\label{statvarlambda} 
Time dependence of  the non-dimensional rate $\varphi(t)$ from DNS, for different
Damk\"ohler numbers. Passive scalar (solid black lines),  parameters corresponding to Fig.~2({\bf a}), green lines, and 
for the parameters corresponding to Fig.~2({\bf b}), violet lines. For all three parameter sets, results of three independent simulations are shown. Also shown is the theoretical estimate $\varphi^\ast=1$ (see main text).}
\end{figure}
\section{Determination of $\varphi(t)$ and $R(t)$}
\label{sec:det}
In this Section, we describe how to determine the unknown rates $\varphi(t)$ and $R(t)$ (see main text). 
We first discuss the function $\lambda(t)$. It describes how the length scale of supersaturation fluctuations decays.
The mapping closure is a one-point approximation (Section \ref{sec:mpc}), so the rate at which the fluctuations relax must be obtained
in another way. Here we extract it from DNS (Section~\ref{sec:dns}). In order to make sure that the model predicts the correct scalar dissipation rate
$\varepsilon_s(t) = 2 \kappa \la \nabla s \cdot \nabla s \ra$,  we must first evaluate
\begin{align}\label{gradsqrone}
 \varepsilon_s(t)= 2\varphi(t) \lp \pder{X}{\eta}\rp^2  \la \pder{\xi (\ve z)}{z_i}  \pder{\xi (\ve z)}{z_i} \ra\,.
\end{align}
This and the following 
equations are written in the non-dimensional units introduced in the main text. 
The average on the r.h.s. is an unconditional average. But since the average conditioned on $\xi=\eta$ in Eq.~(\ref{avgg}) evaluates to unity, independently  of $\eta$, both averages
are the same, conditional and unconditional, and equal to unity.  This implies:
\begin{align}\label{lambda}
\varepsilon_s(t)
= 2\varphi(t) 
{\left( \pder{X}{\eta} \right)^2}\,.
\end{align}
To compute $\varphi(t)$ from  Eq.~(\ref{lambda}) in DNS is straightforward. First, the supersaturation dissipation rate is $\varepsilon_s(t)$ is readily computed from the gradient of $s(\ve x,t)$.  Second, the mapping $X(\eta,t)$ can be computed from the Eulerian cumulative distribution function of supersaturation, using Eq.~(\ref{gausscdf}).
We remark that the DNS results are subject to statistical variability. The evolution of $\varphi(t)$ for three independent DNS at three different parameter combinations is shown Fig.~(\ref{statvarlambda}). Despite the  variability, the trend for $\varphi(t)$ to grow more rapidly at larger Damk\"ohler numbers is evident. The functions $\varphi(t)$  shown in Fig.~2({\bf a}) are averages of the independent functions $\varphi(t)$ in Fig.~\ref{statvarlambda}. Each average 
in Fig.~1({\bf b})  is taken over the corresponding three independent functions in Fig.~\ref{statvarlambda} with identical parameters.

\begin{figure}[b]
\centering
\begin{overpic}[width=140mm]{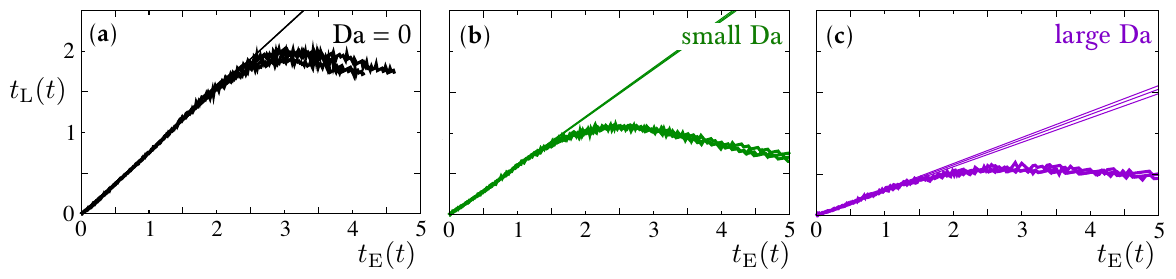}
\end{overpic}
\caption{\label{fig:scaledtimes}
Comparison of the two time scales $t_L(t)$  and $t_E(t)$ extracted from DNS (thick lines).
Also shown are linear fits  in the region  where $t_L(t)$ vs. $t_E(t)$ are roughly proportional (thin lines).
 Results of three independent simulations are shown.
The resulting proportionality constants $C$ (see text) are given in Table~\ref{tab:C}. 
({\bf a}) $\das = 0$.
({\bf b}) Same as ({\bf a}), but at the low Damk\"ohler numbers of Fig.~2({\bf b}). ({\bf c}) Same as ({\bf a}), but
 the large Damk\"ohler numbers of Fig.~2({\bf c}).
}
\end{figure}

Now consider the rate $R(t)$ at which 
the Lagrangian supersaturation distribution  relaxes to the Eulerian one.
We now show how to extract the rate $R(t)$ from DNS, and that it obeys
\begin{align}\label{rfunc}
R(t) = C \varphi(t)
\end{align}
 for not too large times, which allows us to extract the constant $C$.
The rate $R(t)$ is determined so that the Lagrangian supersaturation distribution in the model relaxes in the same way as in the DNS. We measure
the overlap between the two distributions using the Bhattacharyya  overlap \cite{bi2019role}. Computing this measure does not entail dividing with probability densities, whose magnitudes may be small. As a consequence, it  is less sensitive to numerical fluctuations than the perhaps more familiar Kullback-Leibler divergence.
More precisely, we  extract the function 
\begin{align}
t_L(t) = \int_0^t \md t' R(t') \,,
 \end{align} 
from the DNS that minimises the overlap between the Lagrangian  distributions in the statistical model and from the DNS. 
The Lagrangian supersaturation distribution is readily extracted from the DNS. The corresponding distribution in the statistical model is
evaluated as follows. We note that it is given by the distribution  of $\xi(t)$,  because random samples $\xi$ are mapped to the Lagrangian supersaturation distribution by $X(\eta,t)$. Solving the Langevin equation quoted in the main text yields a solution for the distribution of $\xi(t)$ parameterised by $t_L(t)$. Mapping this with $X(\eta,t)$ yields
the model distribution, and we determine $t_L(t)$ so that it minimises the overlap between model and DNS distributions. 
\begin{table}[tb]
\caption{\label{tab:C}
Parameter values for  for the DNS runs,  see Sections~\ref{sec:dns} and~\ref{sec:det} in this supplemental information. 
The large-eddy time $\tau_L$, the droplet evaporation time $\tau_d$, the supersaturation relaxation time $\tau_s$, the Damköhler numbers $\dad$ and $\das$, and the constant $C$ extracted from four indepdent DNS of a passive scalar, four independent DNS at low $\das$t numbers, and three indepdent DNS at high $\das$ numbers.
The DNS in Fig.~2 are for identically evolving velocity fields and droplet positions (see Section~\ref{sec:params}).} 
\begin{tabular}{lccccccccc}
\hline\hline
& $\tau_L [s]$ \hspace*{5mm} & $\taud$ [s] \hspace*{5mm} & $\taus$ [s] \hspace*{5mm} & $\dad$ & $\das$   \hspace*{5mm}& $C$   \\\hline
Fig.~2, passive scalar  & 1.438 & $\infty$ & $\infty$ & 0 & $0$ & 0.78\\
Fig.~2 and 3, low $\das$  & 1.438 & $19.72$ & $1.806$ & 0.073& $0.80$ & 0.62\\
Fig.~2 and 3, high $\das$ & 1.438 & $1.972$ & $0.1806$ & 0.73& $8.0$ & 0.30\\
Figs.~\ref{statvarlambda} and \ref{fig:scaledtimes}, passive scalar &1.479/1.450/1.476& $\infty$ & $\infty$ &0 & $0$ & 0.76/0.76/0.76\\
Figs.~\ref{statvarlambda} and \ref{fig:scaledtimes},  low $\das$&1.463/1.415/1.408&19.72&1.806&$0.074/0.072/0.071$ & $0.81/0.78/0.78$ & 0.60/0.60/0.59\\
Figs.~\ref{statvarlambda} and \ref{fig:scaledtimes},  high $\das$ &1.438/1.435/1.417 &1.972&0.1806& $0.73/0.73/0.72$ & $8.0/7.9/7.8$ & 0.31/0.32/0.30\\\hline\hline
\end{tabular}
\end{table}
The result is shown in Fig.~\ref{fig:scaledtimes}. We plot the Lagrangian time as a function of the Eulerian time
\begin{align}
t_E(t) =
\int_0^t \md t' \varphi(t')
 \end{align}
for the different parameter combinations used in the main text. We see that the two times are proportional to each other initially,
 for at least up to one large-eddy time (not shown).
This implies that Eq.~(\ref{rfunc}) is a good approximation, and allows us to extract the constant $C$ (Table~\ref{tab:C}). The values of   $C$ are subject to some statistical variability, as expected. Nevertheless, a tendency for $C$ to decrease with increasing Damk\"ohler number is evident. This reflects the fact that
phase change suppresses Lagrangian supersaturation fluctuations. 
It is worth noting that it is important to model $R(t)$ accurately at early times, but  it is not very critical at late times, as explained in the main text.

\section{Details regarding Figs.~2 and 3}\label{sec:params}
The parameters of the simulations in Fig.~2({\bf a}) and ({\bf b}) are taken from the DNS study of Ref.~\cite{kumar2012extreme}.
The simulations in Fig.~2({\bf a}) uses parameters from the second row of Table~2 in Ref.~\cite{kumar2012extreme}, and the simulations shown in Fig.~2 ({\bf b}) use parameters from the third row of that table. We extracted non-dimensional parameters from Ref.~\cite{kumar2012extreme} as follows.
We extracted the  turbulent
dissipation rate $\varepsilon$, and used that to force the velocity field as described in Section~\ref{sec:dns}. 
We then determined the stationary turbulent kinetic energy $k$ in our DNS  by averaging the instantaneous kinetic energy over the duration of the simulation. 
This yields the eddy-turnover time, $\tau_L = k/\varepsilon$. The values obtained are summarised in Table~\ref{tab:C}.
To compute the Damk\"ohler numbers $\dad$ and $\das$, we 
 extracted from Ref.~\cite{kumar2012extreme} the initial droplet radius $r_0 = 20 \, \mu\text{m}$, the droplet-number density $n_0$ = 164~cm$^{-3}$ of the initially cloudy air, the density $\varrho_\text{w} = 1000$~kg/m$^3$ of liquid water, the saturation water-vapour mixing ratio $q_{\text{vs}} = 0.00356$, and the mass density $\airdensity = 1.06$~kg/m$^3$ of dry air. Comparing Eq.~(1b) to Eq.~(3) of Ref.~\cite{kumar2012extreme}, we inferred $A_2 = 1/(\airdensity q_{\text{vs}}) = 265$~m$^3$/kg. Then we extracted the supersaturation $s_\text{e} = -0.2$ of the initially dry air from Fig.~1 of Ref.~\cite{kumar2012extreme}. 
This allowed us to compute $\tau_\text{s} = (4\pi A_2 A_3 \varrho_w n_0 r_0 )^{-1} = 1.806$~s and $0.1806$~s, as well as 
$\taud = r_0^2/(2 A_3|s_{\rm e}|) = 19.72$~s and $1.972$~s, for the second and third row of Table~2 of Ref.~\cite{kumar2012extreme}, using the values $A_3 = 5.07\times 10^{-11}$~m$^2$/sand $5.07\times 10^{-10}$~m$^2$/s given there. The resulting Damk\"ohler numbers are quoted in Table~\ref{tab:C}. Their values differ slightly from the corresponding values quoted in Ref.~\cite{kumar2012extreme}. This reflects that statistically independent DNS with identical values of $\tau_\text{s}$ and $\taud$, and $\varepsilon$ may have slightly different turbulent kinetic energies $k$, leading to different values of $\tau_L$, $\dad$ and $\das$. In order to study the effects of droplet phase change independently of the statistical variability of DNS, all DNS reported in Figs.~2 and~3 are shown for identically evolving velocity fields and droplet positions.

The initial mapping was computed using Eq.~(\ref{gausscdf}) and the initial supersaturation profile in Fig.~1 of Ref.~\cite{kumar2012extreme}. From that figure, we extracted the initial volume fraction $\chi = 0.4$ of cloudy air and used this volume fraction to initialise $\xi(t)$-values for the droplets. To this end, we sampled $\xi(t=0)$ from a standardised Gaussian distribution conditional on $\xi(t) > \Phi^{-1}(1-\chi)$, where $\Phi^{-1}$ was the inverse cumulative distribution function of a standardised normal distribution. Eq.~(\ref{gausscdf}) implies that this initialisation corresponds to a uniform distribution of droplets in  the fraction $\chi$ of air with the largest supersaturation
[Fig.~1({\bf a})].

\section{Details regarding Fig.~3}
The conditional average $\langle n(t)|s(t)=s\rangle$ in Figure~3 was obtained  as follows from the DNS data:
\begin{equation}
\langle n(t)|s(t)\!=\!s\rangle=\int{\rm d}n' n' \mathscr{P}(n'|s)\,.
\end{equation}
The conditional probability was computed as $\mathscr{P}(n|s) = \mathscr{P}(n,s)/\mathscr{P}(s)$, where $\mathscr{P}(n,s)$ is
the joint distribution of droplet number density $n$ and supersaturation $s$, and $\mathscr{P}(s)$ is the distribution supersaturation, both obtained from the DNS.
Since the average droplet-number density  tended to be quite small, we found it necessary to use non-uniform bins in the droplet number $n$ to represent  $\mathscr{P}(n',s)$,
choosing smaller bins for smaller droplet number densities, to resolve the shape of the distribution at small values of $n$. To this end, we used half-interval Chebyshev nodes.

\section{Model prediction for droplet-size distribution}
 
\begin{figure}
\centering
\begin{overpic}[width=60mm]{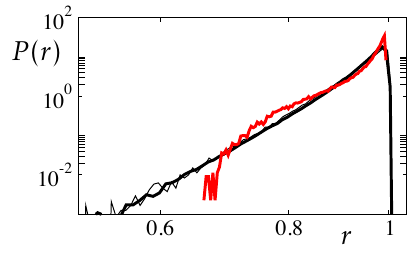}
\end{overpic}
\vspace{3mm}
\caption{\label{fig:large_moist}
 Steady-state droplet-size distribution $P(r)$ versus droplet radius $r$. Present model (thick black line), DNS results
 for the same parameters (thin black line), and model from Ref.~\cite{fries2021key} (red line). Parameters from Ref.~\cite{kumar2012extreme}: $\das = 8.0$, Damk\"ohler ratio $\mathscr{R}=\dad/\das= 0.09$. 
}
\end{figure}
Fig.~\ref{fig:large_moist} shows droplet-size distributions for parameters values extracted from  Kumar {\em et al.}~\cite{kumar2012extreme}.
Shown are DNS results (thin black line), the results of the present model (thick black line), and results of the  earlier statistical model of Fries {\em et al.}~\cite{fries2021key}
(red line). Both models are in qualitative
agreement with the DNS results, but the new model agrees better with the DNS in the tails, compared with the model from Ref.~\cite{fries2021key}. 
The latter describes mixing of supersaturation imposing exponential relaxation to a mean field. This ensures that supersaturation remains bounded, as it should, but the model from Ref.~\cite{fries2021key} does not account for correlations between $n$ and $s$, unlike the present model.  As a consequence, the model from Ref.~\cite{fries2021key} fails
to describe the most rapidly evaporating droplets. 


%